\documentclass[prd,reprint,showpacs,showkeys]{revtex4-1}
\usepackage{amsfonts}
\usepackage{mdframed}
\usepackage{amssymb}
\usepackage{amsmath}
\usepackage{graphicx}
\usepackage[font={footnotesize,it}]{caption}

\begin{document}

\preprint{}
\title{Screening of the Reissner-Nordstr\"{o}m charge by a thin-shell of
dust matter}
\author{S. Habib Mazharimousavi}
\email{habib.mazhari@emu.edu.tr}
\author{M. Halilsoy}
\email{mustafa.halilsoy@emu.edu.tr}
\affiliation{Department of Physics, Eastern Mediterranean University, Gazima\u{g}usa,
North Cyprus, Mersin 10 - Turkey.}

\begin{abstract}
A concentric charged thin-shell encircling a Reissner-Nordstr\"{o}m black
hole screens the clectric / magnetic charge completely to match with an
external Schwarzschild black hole. The negative mass thin-shell is shown to
be stable against radial perturbations. It is shown further that by
reversing the roles of inside Reissner-Nordstr\"{o}m and outside
Schwarzschild geometries the mass of the appropriate shell becomes positive.
\end{abstract}

\maketitle

\section{Introduction}

Black holes are highly localized simplest objects in our cosmos that may
carry charges (i.e. hairs) of various kinds. No-hair conjecture \cite{NHC}
refers simply to the degrees of freedom other than the well-accepted ones
such as mass, electric / magnetic charge and the angular momentum. Internal
degrees of freedom such as the non-abelian gauge charges can naturally be
added to the abelian electromagnetic charges to extend the list of hairs for
the black holes.

With the advent of surface-layer formalism and thin-shells in general
relativity \cite{1,2,3,4,5,6,7}, the question naturally arises whether the
hairs of the black hole can be screened against external observer at
infinity. Thin-shell and its stability in Schwarzschild black hole spacetime
was studied by Brady, Louko and Poisson in \cite{BLP} where they have shown
that a thin-shell with positive energy density which satisfies the dominant
energy conditions may be stable against a radial perturbation. Nonradial
linear oscillations of shells was studied by Schmidt in \cite{Schmidt} and
inclusion of a cosmological constant was done by Ishak and Lake in \cite{IL}%
. A generalized study on thin-shells in vacuum was investigated by Goncalves 
\cite{Goncalves} and following \cite{BLP}, Lobo and Crawford \cite{LC}, by
considering a spacetime satisfying the transparency condition, have studied
generic dynamic spherically symmetric thin-shells. Accelerated shells and
the relation between stress-energy and motion for such layers has been
considered by Krisch and Glass in \cite{KG} while the stability of charged
thin-shells has been studied by Eiroa and Simeone \cite{ES} and the same
authors worked on thin-shells in $2+1-$dimensions including Born-Infeld
matter sources in the bulk \cite{ES2}. Very recently stability of thin-shell
interfaces inside compact stars has been worked out by Pereira, Coelho, and
Rueda in \cite{PCR}. Application of thin-shell formalism in making dark
energy stars has been considered by Bhara and Rahaman in \cite{BR} while the
gravitational vacuum star or \textit{gravastar,} based on the same
formalism, has been proposed by Mazur and Mottola in \cite{MM} which was
developed further by Visser and Wiltshire in \cite{VW} and the references
therein.

Our aim in this study is to apply such a formalism to the standard
Reissner-Nordstr\"{o}m (RN) black hole which carries a static electric or
magnetic charge. The thin-shell is assumed concentric with the RN black hole
and with a radius greater than the event horizon of the latter. A toy-model
version of our formalism in $2+1-$dimensional case was considered before in
connection with the regular Bardeen black hole \cite{9}. The source of the
Bardeen black hole therein was assumed to be of non-linear electrodynamic
origin. In the present problem the inside spacetime is taken to be RN while
the external spacetime is Schwarzschild geometry. Application of the
boundary conditions at the thin-shell in between serves to screen the charge
of the internal RN geometry against outside. That is, beyond the thin-shell
no trace of electric charge $Q$ of the RN black hole is left. In this sense
the thin-shell acts as a perfect absorber of the electric charge of the RN
black hole. Such a screening process, however, is not without consequences.
The imposed boundary conditions fix the mass / energy and charge of the
shell to play its absorbent role. The energy density $\sigma $ of the
thin-shell at equilibrium radius turns out to be negative, $\sigma <0.$ This
situation is known to be notorious enough in the topic of wormholes and
thin-shell wormholes \cite{10,11,12,14,15,16}, although it may be considered
natural in the realm of quantum field theory. Once this situation is taken
for granted we proceed with the stability analysis of the thin-shell against
linear radial perturbations. With perturbation, besides the energy density
at the equilibrium state we have emerging pressure tensions satisfying an
equation of state of the form $\frac{dP}{d\sigma }=\omega ,$ where $P=$%
pressure, $\sigma =$energy density and $\omega $ is a constant proportional
to the speed of sound. It turns out that the thin-shell around a RN black
hole which screens its electric / magnetic charge from outside is stable
against radial perturbations. The stability configuration is plotted
numerically. We may anticipate that a similar analysis can naturally be
carried out for a black hole carrying a Yang-Mills charge. Since these
non-abelian gauge charges are trapped / confined inside nuclei it may lead
effectively to a geometrical theory of confinement for fractionally charged
fermions. For this purpose, however, the relevant proper boundary conditions
should be those of Einstein-Maxwell-Dirac-Yang-Mills theory, which lies
beyond our scope in this paper. Let us add that the range of applications
for our method seems limitless and for all these, thanks to the Einstein's
junction equations with tuned sources satisfied on layers / surfaces.

\section{The Formalism}

Let's assume that a RN black hole with mass $m$ and total charge $Q$ sits at
the origin of a spherically symmetric spacetime whose event horizon is
located at $r=r_{e}.$ A timelike thin-shell of dust is located at $r=a>r_{e}$
with energy momentum tensor $S^{ij}=diag\left[ \sigma ,0,0\right] $ with
respect to an observer on the shell of line element 
\begin{equation}
ds^{2}=-d\tau ^{2}+a^{2}\left( \tau \right) d\Omega ^{2}
\end{equation}%
in which $\tau $ is the proper time on the shell and $d\Omega ^{2}$ is the
line element on $S^{2}$. Next, we are interested to see the possibility of
having charge of the black hole unseen by a distant observer. In other
words, is it possible to have the spacetime outside the shell a
Schwarzschild black hole with mass $M,$ different from $m,$ but related to $%
m $ and $Q?$

Using the well-known Darmois-Israel formalism \cite{1,2} one may consider
two pseudo Riemannian manifolds $\mathcal{M}_{1}$ and $\mathcal{M}_{2}$ with
identical timelike boundaries located at $r=a$ with spherically symmetric
line elements 
\begin{multline}
ds_{1}^{2}=-f_{1}\left( r\right) dt^{2}+\frac{1}{f_{1}\left( r\right) }%
dr^{2}+r^{2}d\Omega ^{2},\text{ } \\
\text{for }r\leq a\text{ and }f_{1}\left( r\right) =1-\frac{2m}{r}+\frac{%
Q^{2}}{r^{2}}
\end{multline}%
and 
\begin{multline}
ds_{2}^{2}=-f_{2}\left( r\right) dt^{2}+\frac{1}{f_{2}\left( r\right) }%
dr^{2}+r^{2}d\Omega ^{2}, \\
\text{ for }r\geq a\text{ and }f_{2}\left( r\right) =1-\frac{2M}{r}
\end{multline}%
in which $m$ and $Q$ are the mass and charge of the inner black hole while $M
$ is a constant to be identified. By gluing these manifolds from their
boundaries we construct a complete manifold $\mathcal{M}$. The hypersurface
boundaries used for gluing is given by (1) and the Israel junction
conditions imposes ($c=G=1$)%
\begin{equation}
k_{i}^{j}-k\delta _{i}^{j}=-8\pi S_{i}^{j}.
\end{equation}%
Here $k_{i}^{j}=K_{i}^{j\left( 2\right) }-K_{i}^{j\left( 1\right) }$ is the
effective extrinsic curvature tensor of the shell with $K_{i}^{j\left(
2\right) }$ and $K_{i}^{j\left( 1\right) }$ on each side of the shell and $%
k=k_{i}^{i}$ is the extrinsic curvature scalar of the shell. In brief 
\begin{equation}
K_{ij}^{\left( 1,2\right) }=-n_{\gamma }^{\left( 1,2\right) }\left( \frac{%
\partial ^{2}x^{\gamma }}{\partial y^{i}\partial y^{j}}+\Gamma _{\alpha
\beta }^{\gamma }\frac{\partial x^{\alpha }}{\partial y^{i}}\frac{\partial
x^{\beta }}{\partial y^{j}}\right) 
\end{equation}%
with 
\begin{equation}
n_{\gamma }^{\left( 1,2\right) }=\frac{\partial _{\gamma }F}{\sqrt{g^{\alpha
\beta \left( 1,2\right) }\partial _{\alpha }F\partial _{\beta }F}}
\end{equation}%
the normal 4-vector on the sides of the shell given by the surface $F=r-a=0$%
. \ We note that $x^{\alpha }\in \left\{ t,r,\theta ,\varphi \right\} $
while $y^{i}\in \left\{ \tau ,\theta ,\varphi \right\} $ and $\partial
_{\alpha }=\frac{\partial }{\partial x^{\alpha }}.$ For the static
thin-shell, the explicit calculation admits%
\begin{equation}
\frac{1}{4\pi a}\left( \sqrt{f_{1}}-\sqrt{f_{2}}\right) =\sigma 
\end{equation}%
and%
\begin{equation}
\frac{2f_{2}+af_{2}^{\prime }}{16\pi a\sqrt{f_{2}}}-\frac{%
2f_{1}+af_{1}^{\prime }}{16\pi a\sqrt{f_{1}}}=0
\end{equation}%
in which a prime stands for the derivative with respect to $r$ and all
functions are calculated at $r=a.$ In order to have the second condition
satisfied we must have 
\begin{equation}
M=\frac{a\left( r_{e}-m\right) }{\left( a-m\right) +\left( r_{e}-m\right) }
\end{equation}%
in which $r_{e}$ is the event horizon of the RN black hole given by%
\begin{equation}
r_{e}=m+\sqrt{m^{2}-Q^{2}}.
\end{equation}%
Having $M,$ one can find $\sigma $ of the thin-shell which is given by%
\begin{multline}
\sigma =-\frac{1}{4\pi a}\left( \frac{\sqrt{a-r_{e}}}{\sqrt{\left(
a-m\right) +\left( r_{e}-m\right) }}\right. - \\
\left. \frac{\sqrt{a\left( a-2m\right) -r_{e}\left( r_{e}-2m\right) }}{a}%
\right) .
\end{multline}%
Herein, due to the fact that $a>r_{e}$ and $m\leq r_{e}\leq 2m,$ $\sigma $
remains real but negative. Therefore the shell is like a bubble of exotic
matter. The amount of total exotic matter can be found as%
\begin{multline}
\Omega =\int\nolimits_{0}^{2\pi }\int\nolimits_{0}^{\pi
}\int\nolimits_{0}^{\infty }\sigma \delta \left( r-a\right) \sqrt{-g}%
drd\theta d\varphi =4\pi a^{2}\sigma = \\
-\left( \frac{a\sqrt{a-r_{e}}}{\sqrt{\left( a-m\right) +\left(
r_{e}-m\right) }}-\sqrt{a\left( a-2m\right) -r_{e}\left( r_{e}-2m\right) }%
\right) .
\end{multline}%
In terms of $m,$ $M$ and $Q$ we find%
\begin{equation}
\Omega =-\sqrt{\left( m-\Delta \right) \left( m+\Delta -2M\right) },
\end{equation}%
in which $\Delta =\sqrt{m^{2}-Q^{2}}.$ Let's remark that at the extremal
limits one finds%
\begin{equation}
\lim_{Q\rightarrow 0}M=m,\text{ }\lim_{Q\rightarrow 0}\Omega =0,\text{ }
\end{equation}%
which implies the absence of the thin-shell and 
\begin{equation}
\lim_{Q\rightarrow m}M=0,\lim_{Q\rightarrow m}\Omega =-m=-Q.\text{ }
\end{equation}%
Here $\sigma =-\frac{m}{4\pi a^{2}}=-\frac{Q}{4\pi a^{2}}$ which is nothing
but the mass density over the surface area of the thin-shell with the same
mass and charge of the black hole. It should be added that the charge of the
spherical thin-shell must be the negative ($-Q$) of the black hole charge.
This can be verified as follows.

The electric potential of the Reissner-Nordstr\"{o}m black hole, up to a
gauge transformation can be expressed appropriately by 
\begin{equation}
\mathbf{A}=Q\left( \frac{1}{r}-\frac{1}{a}\right) \Theta \left( a-r\right)
dt.
\end{equation}%
Note that $\Theta \left( a-r\right) $ is the unit step function defined by 
\begin{equation}
\Theta \left( a-r\right) =\left\{ 
\begin{array}{cc}
1, & r<a \\ 
0, & r>a%
\end{array}%
\right. .
\end{equation}%
Accordingly, the Maxwell $2-$form is 
\begin{equation}
\mathbf{F}=\frac{Q}{r^{2}}\Theta \left( a-r\right) dt\wedge dr
\end{equation}%
with its dual $2-$form%
\begin{equation}
^{\star }\mathbf{F}=Q\sin \theta \Theta \left( a-r\right) d\theta \wedge
d\varphi .
\end{equation}%
The sourceful Maxwell equation takes the form%
\begin{equation}
d\left( ^{\star }\mathbf{F}\right) =\text{ }^{\star }\mathbf{j}
\end{equation}%
with the current density $3-$form%
\begin{equation}
^{\star }\mathbf{j}=-Q\sin \theta \delta \left( a-r\right) dr\wedge d\theta
\wedge d\varphi
\end{equation}%
in which $\delta \left( a-r\right) $ stands for the Dirac delta function on
the shell.

The integral of $^{\star }\mathbf{j}$ yields the charge on the shell as 
\begin{equation}
-Q=\frac{1}{4\pi }\int \text{ }^{\star }\mathbf{j}.
\end{equation}%
This verifies that beyond the spherical thin-shell charge does not exist,
justifying the Schwarzschild metric.

To conclude this section we would like to express our variables and
quantities in terms of $m.$ By introducing $\frac{a}{m}=\alpha ,$ and $\frac{%
Q^{2}}{m^{2}}=\epsilon $ we find%
\begin{equation}
\frac{M}{m}=\frac{\alpha \sqrt{1-\epsilon }}{\alpha -1+\sqrt{1-\epsilon }},
\end{equation}%
\begin{equation}
\frac{\Omega }{m}=-\frac{\left( 1-\sqrt{1-\epsilon }\right) \sqrt{\alpha
^{2}-2\alpha +\epsilon }}{\alpha -1+\sqrt{1-\epsilon }}
\end{equation}%
and%
\begin{equation}
\sigma m=-\frac{\sqrt{\alpha ^{2}-2\alpha +\epsilon }}{4\pi \alpha ^{2}}%
\frac{1-\sqrt{1-\epsilon }}{\alpha -1+\sqrt{1-\epsilon }},
\end{equation}%
with $0\leq \epsilon \leq 1$ and $1+\sqrt{1-\epsilon }<\alpha .$ These allow
us to set the quantity $m$ to unity without losing the generality of the
problem. In the same line we remark that the location of horizon (i.e., the
horizon observed by a distant frame) is given by 
\begin{equation}
\bar{r}_{e}=2M=m\frac{2\alpha \sqrt{1-\epsilon }}{\alpha -1+\sqrt{1-\epsilon 
}}.
\end{equation}%
The latter clearly shows that $\frac{\bar{r}_{e}}{m}=\frac{2\alpha \sqrt{%
1-\epsilon }}{\alpha -1+\sqrt{1-\epsilon }}<\alpha $ which implies that the
event horizon is located inside the shell with respect to a distant observer.

\section{Stability analysis}

As we have shown in the previous section, one can consider a thin-shell of
exotic dust surrounding a RN black hole which screens the electric charge of
the black hole. The resulting solution from a distant observer will be a
Schwarzschild black hole with a new effective mass. In this section we shall
investigate the stability of such a thin-shell. To do so let's consider the
radius of the shell to be a function of $\tau $ the proper time on the
shell. By using the Israel formalism we find 
\begin{equation}
\sigma =\frac{1}{4\pi a}\left( \sqrt{f_{1}+\dot{a}^{2}}-\sqrt{f_{2}+\dot{a}%
^{2}}\right)
\end{equation}%
and%
\begin{equation}
P=-\frac{\sigma }{2}+\frac{1}{16\pi }\left( \frac{2\ddot{a}+f_{2}^{\prime }}{%
\sqrt{f_{2}+\dot{a}^{2}}}-\frac{2\ddot{a}+f_{1}^{\prime }}{\sqrt{f_{1}+\dot{a%
}^{2}}}\right) .
\end{equation}%
We note that although for the static equilibrium we assumed $P=0$ in the
dynamic regime, $P$ may not be zero as the matter is not at rest any more.
Therefore while we consider the relation between $M$ and other parameters
given by (9) i.e., dictated by static equilibrium, in the dynamic case we
adopt $\frac{dP}{d\sigma }=\omega $ in which $\omega $ is a real constant.
The energy conservation, on the other hand, imposes%
\begin{equation}
S_{;j}^{ij}=0
\end{equation}%
in which for $i=\tau $ one finds%
\begin{equation}
\sigma ^{\prime }=-\frac{2}{a}\left( \sigma +P\right)
\end{equation}%
with $^{^{\prime }}\equiv \frac{d}{da}.$

%
\begin{figure}[h]
\includegraphics[width=80mm,scale=0.7]{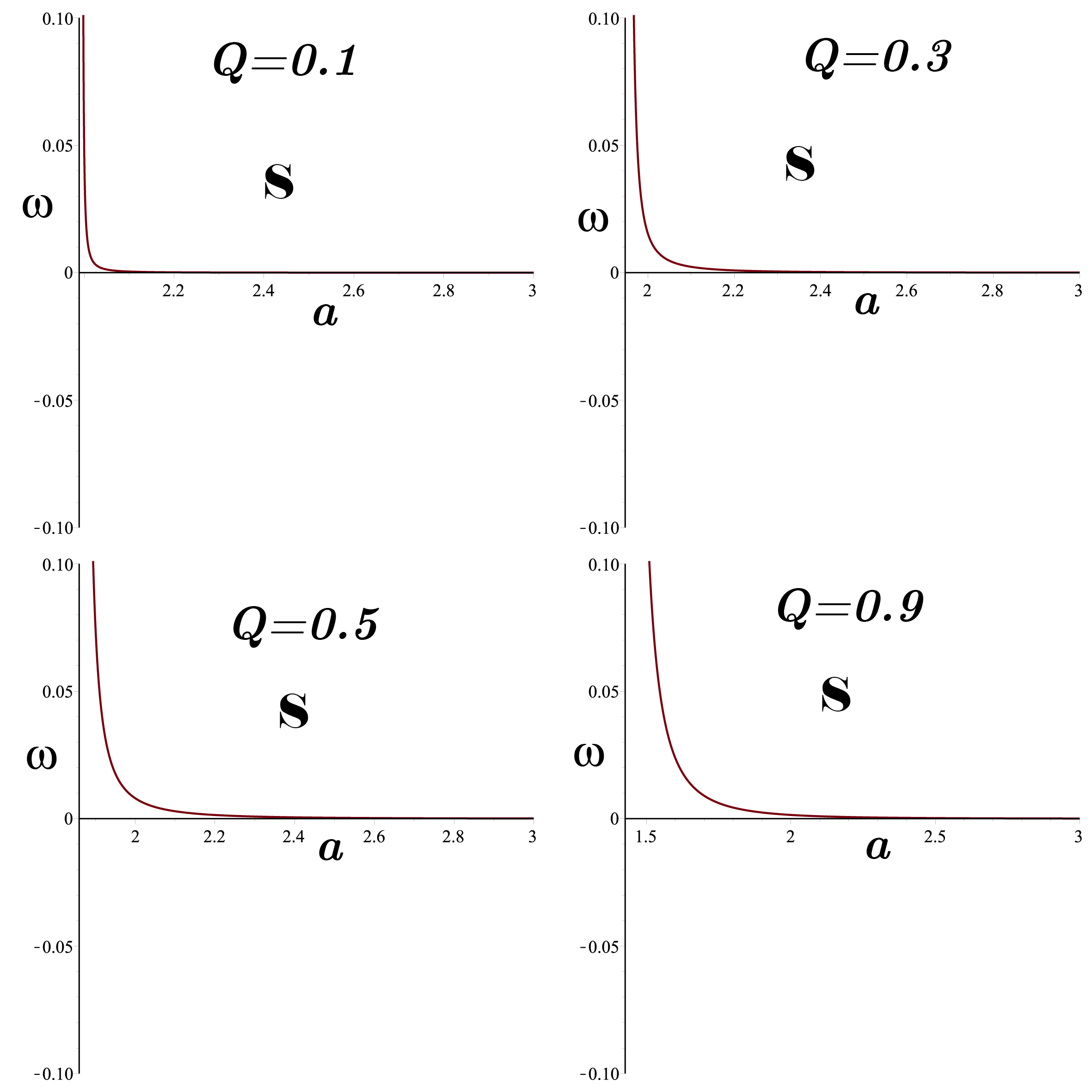}
\caption{A plot of $V^{\prime \prime }=0$ at $a=a_{0}$ in terms of $%
a(=a_{0}) $ and $\protect\omega $ for $m=1.0$ and $Q=0.1,0.3,0.5,0.9.$ The
region with $V^{\prime \prime }>0$ which is the stability zone is also
indicated by \textbf{S}. The effect of charge on stability / instability is
clearly seen. }
\end{figure}

As a result from Eq. (27) we find a one dimensional equation of motion for
the dynamical shell given by \cite{ES}%
\begin{equation}
\dot{a}^{2}+V\left( a\right) =0
\end{equation}%
in which 
\begin{equation}
V\left( a\right) =-4\pi ^{2}a^{2}\sigma ^{2}+\frac{f_{1}+f_{2}}{2}-\frac{%
\left( f_{1}-f_{2}\right) ^{2}}{64\pi ^{2}a^{2}\sigma ^{2}}.
\end{equation}%
We must add that in (31), $\sigma =\sigma \left( a\right) $ is a function of 
$a$ - the radius of the shell after perturbation - which is no longer the
same as its equilibrium value, say $\sigma _{0}=\sigma \left( a=a_{0}\right) 
$ which is given by%
\begin{equation}
\sigma _{0}=\frac{1}{4\pi a_{0}}\left. \left( \sqrt{f_{1}}-\sqrt{f_{2}}%
\right) \right\vert _{a=a_{0}}.
\end{equation}%
In order to have the thin-shell stable against a radial perturbation, Eq.
(32) must admit an oscillatory motion which means that at the equilibrium
point (say at $a=a_{0}$) where $V\left( a_{0}\right) =$ $V^{\prime }\left(
a_{0}\right) =0$ and $V^{\prime \prime }\left( a_{0}\right) >0.$ In Fig. 1
we plot $V^{\prime \prime }\left( a_{0}\right) $ for the specific value of $%
m=1.0$ and $Q=0.2.$ As one observes in the region with $\omega >0$ the
thin-shell is stable while otherwise it occurs for $\omega <0.$ Let's add
also that, $\omega =cons.$ yields%
\begin{equation}
P=\omega \left( \sigma -\sigma _{0}\right) \text{ }
\end{equation}%
which in turn implies that with $\sigma >\sigma _{0}$ / $\sigma <\sigma _{0}$
and $\omega >0,$ to have the shell stable, the pressure must be negative /
positive after the perturbation. We add that this behavior is not only for
the specific value of $m$ and $Q.$

As a final remark we comment that our formalism allows to interchange the
roles of inner and outer spacetimes. That means this time the inner
spacetime is Schwarzschild while the outer one is the RN. We have%
\begin{multline}
ds_{1}^{2}=-f_{1}\left( r\right) dt^{2}+\frac{1}{f_{1}\left( r\right) }%
dr^{2}+r^{2}d\Omega ^{2},\text{ } \\
\text{for }r\leq a\text{ and }f_{1}\left( r\right) =1-\frac{2M}{r}
\end{multline}%
and 
\begin{multline}
ds_{2}^{2}=-f_{2}\left( r\right) dt^{2}+\frac{1}{f_{2}\left( r\right) }%
dr^{2}+r^{2}d\Omega ^{2},\text{ } \\
\text{for }r\geq a\text{ and }f_{2}\left( r\right) =1-\frac{2m}{r}+\frac{%
Q^{2}}{r^{2}}.
\end{multline}%
Obviously the thin-shell of radius $r=a$ carries the charge $Q$ to make the
charge of the external RN geometry. Following the foregoing analysis we
conclude that $M$, $m$ and $Q$ are related as given in Eq. (22) while the
energy density is positive now given by%
\begin{equation}
\sigma =\frac{\sqrt{\alpha ^{2}-2\alpha +\epsilon }}{4\pi m\alpha ^{2}}\frac{%
1-\sqrt{1-\epsilon }}{\alpha -1+\sqrt{1-\epsilon }}
\end{equation}%
in which $\epsilon $ and $\alpha $ are as before and consequently $\Omega
>0. $ In this case the charge distribution lies on the spherical shell at $%
r=a$. The new thin-shell is also stable against a radial perturbation with
an equation of state $\frac{dP}{d\sigma }=\omega >0$ as in the other case.

\section{Conclusion}

Although in the present study we investigated the erasure of a RN black hole
charge to the external world through junction conditions imposed on a
contrived outer thin-shell the method seems more generic, apt for more
general central objects. We have shown, for instance, that the roles of
inner RN and outer Schwarzschild geometries is reversible with a positive
mass on the shell. Not to mention, a counter-rotating cylindrical shell may
absorb the rotational hair of a black hole to turn it into a static one. The
question may be raised: can naturally formed absorber shells, thin or thick
in cosmology hide / screen the reality from our telescopes? If yes, then the
effect of screening becomes as important as the lensing of light while
passing near massive heavenly objects. No doubt this may revise our ideas of
black holes and their no-hair theorem. More interestingly this may pave the
way toward a geometrical description of quark confinement provided the
boundary conditions are modified to cover the Dirac and Yang-Mills fields.
Naturally this takes us away from classical physics into the realm of
gravity coupled QCD. Let us add that hiding of charge by geometrical
structures has been considered before for example in \cite{Guen}. Therein
with cited references, it has been shown that non-linear contributions (see
for instance \cite{Guen2}) in an effective theory beyond standard
Einstein-Maxwell plays crucial roles. Finally we must admit that the
negative mass of the thin, stable layer encountered in the formalism remains
to be our concern.

It should also be supplemented that the extremal RN case with $m=Q\left(
\epsilon =1\right) $ with the flat space ($M=0$) inside constitutes a
particular case. The energy density of the thin-shell which becomes a bubble
now takes the form%
\begin{equation}
\sigma =\frac{1}{4\pi m\alpha ^{2}},\text{ (}\alpha >1\text{).}
\end{equation}%
From a different approach the similar problem was considered also in \cite%
{metin}. It follows that such a bubble in an extremal RN spacetime becomes
stable against perturbations described above. As a final remark let us add
that our results have holographycal implications which may be investigated
further.

\bigskip

\end{document}